\documentclass[12pt]{iopart}
\usepackage{graphicx}\usepackage{amssymb,bm}
\usepackage{showlabels,showtags}
\usepackage{verbatim}
\newcommand{\beq}{\begin{equation}}
\newcommand{\eeq}{\end{equation}}
\begin{document}
\title{Electric field formulation for thin film magnetization problems}
\author{John W Barrett$^{1}$ and Leonid Prigozhin$^2$}
\address{$^{1}$Dept.\ of
Mathematics, Imperial College London, London SW7 2AZ, UK.}
\ead{j.barrett@imperial.ac.uk}

\vspace{2mm}

\address{$^2$Dept. of Solar Energy and Environmental Physics,
Blaustein Institutes for Desert Research,
Ben-Gurion University of
the Negev, Sede Boqer Campus, 84990 Israel.}
\ead{leonid@math.bgu.ac.il}

\begin{abstract}
We derive a variational formulation for thin film magnetization problems in type-II
superconductors written in terms of two variables, the electric field and the
magnetization function. A numerical method, based on this formulation, makes it possible
to accurately compute all variables of interest, including the electric field,
for any value of the power in the power law current-voltage relation characterizing the
superconducting material. For high power values we obtain a good approximation to the
critical state model solution. Numerical simulation results are presented for
simply and multiply connected films, and also for an inhomogeneous film.
\end{abstract}
\maketitle

\section{Introduction}
A thin superconductor film in a perpendicular magnetic field is the configuration typical
of experiments with superconducting materials, and is employed in various physical devices
(SQUIDs, magnetic traps for cold atoms, etc.). Macroscopically, the magnetization of
type-II superconductors is well described by eddy current models with critical state
\cite{Bean,Kim} or power law \cite{Rhyner} current-voltage relations.
Solving these highly nonlinear eddy current problems helps to understand the
peculiarities of magnetic flux penetration into thin films,
and is necessary for the design of superconductor-based electronic devices.

Analytically, the sheet current density is known for
the Bean critical state model in both the thin disk \cite{MikhKuz,ClemSanchez} and strip
\cite{BrandtIndenbomForkl} geometries. Numerical methods for modeling magnetization
in flat films of arbitrary shapes were derived, for the power law model,
by Brandt and co-workers in \cite{Brandt95,SchusterB96}; 
see also \cite{VSGJ07,VSGJ08,VSGJ12} and the references therein.
For the critical state model
a numerical scheme, based on a variational formulation of the thin film magnetization
problems, has been proposed in \cite{P98}; see also \cite{NSDVCh}.
Common to these numerical algorithms is the use
of a scalar magnetization (stream) function as the main variable. The sheet current density
in the film is obtained as the 2D curl of this function;
the magnetic field can then be computed from the current density via the Biot-Savart law
and compared to magneto-optical imaging results.

The electric field in a superconductor is also of much interest: it is needed to find the
distribution of the energy dissipation, which is often very nonuniform and can cause
thermal instabilities. Computing the electric field $\bi{e}$ by means of existing numerical
schemes can, however, be difficult for the power law model,
\beq \bi{e}=e_0(j/j_{\rm c})^{p-1}\bi{j}/j_{\rm c}, \label{power}\eeq
where $\bi{j}$ is the sheet current density, $j=|\bi{j}|$, $e_0$ is a constant, $j_{\rm c}$
is the critical sheet current density, and the power $p$ is, typically, between 10 and 100.
Indeed, even if the magnetization function is found with good accuracy, its numerical
derivatives determining the sheet current density in the film are, inevitably,
much less accurate. Computing the electric field via the constitutive relation (\ref{power})
increases the error further and makes it unacceptably large if the power $p$ is high.

As is well-known, the critical state model current-voltage relation can be regarded as
the $p\rightarrow \infty$ limit of the power law (\ref{power}); see \cite{BP00}
for a rigorous proof. The limit can be described as
\beq |\bi{j}|\leq j_{\rm c};\quad \mbox{if }|\bi{j}|< j_{\rm c}\mbox{ then }
\bi{e}=\bf{0};\quad\mbox{if }\bi{e}\neq \bf{0}\mbox{ then } \bi{e}\,\|\,\bi{j}\label{crit}.\eeq
The electric field in this model can be nonzero only in a region where
the current density is critical; there the field is parallel to current density and is
determined by the eddy current problem
with the constitutive relation (\ref{crit}). Note that even if the current density was computed,
e.g., by means of the numerical scheme \cite{P98}, the multi-valued relation (\ref{crit})
alone is not sufficient for the reconstruction of the electric field.
Approximating the electric field in a critical state model is relatively straightforward only
for an infinite strip or a long superconducting cylinder in a perpendicular field \cite{P11}.

For cylinders of an arbitrary cross-section in a parallel field, the magnetic field in
the superconductor can be expressed via the distance to the boundary function
(see \cite{BP10}) or, in more complicated cases, found numerically.
The current density is computed as the 2D curl of this field.
Computing the electric field, however, remains non-trivial. A numerical algorithm for the
electric field reconstruction, requiring integration along the paths of the magnetic flux
penetration, has been proposed in \cite{BL}. A dual/mixed variational formulation of
magnetization problems served as a basis for the efficient computation of the electric field
in \cite{BP06,BP10}.

Determining the electric field in thin film problems is more difficult. Under the simplifying
assumption that the time derivative of the normal to the film magnetic field is,
in the flux penetrated region, close to the ramping rate of the external field,
approximate analytical expressions
for the electric field were found for
the Bean critical state model for the
rectangular and related film shapes in \cite{SchusterB96}.

Here we extend the approach
\cite{BP06,BP10} and derive for thin film magnetization problems a convenient mixed variational
formulation in terms of two variables: the electric field and a scalar auxiliary variable
analogous to the magnetization function.
We use Raviart-Thomas elements of the lowest order \cite{Carst} to approximate
the (rotated) electric field
and  a continuous piecewise linear approximation for the auxiliary
variable. Based on this approximation of the variational problem, our iterative numerical
algorithm suffers no accuracy loss of the computed electric field even for very high values of
the power $p$ in (\ref{power}). Hence, the algorithm can be used to find the electric and
magnetic fields, and the current density for both the power and critical state model
problems.

In this work we focus on the derivation of the mixed variational
formulation, describe the numerical algorithm, and present simulation results.
Rigorous mathematical arguments, including the exact function space set up, and a proof of the
algorithm convergence, etc., will be presented elsewhere \cite{BPmath}.

\section{Magnetization model: a mixed formulation}
Let $\Omega\subset \mathbb{R}^2$ be a domain
and, in the infinitely thin approximation, the superconducting film occupies
the set $\{\overline{\Omega}\times 0\}\subset \mathbb{R}^3$. By $\bi{e}_{\rm i}(x_1,x_2,t)$,
where $(x_1,x_2)\in \Omega$, we denote the tangential to the film (and continuous on it)
component of the electric field $\bi{e}(x_1,x_2,x_3,t)$, and assume it is related to
the film sheet current density $\bi{j}(x_1,x_2,t)$
by the power law (\ref{power}). This law can be re-written as
\beq \bi{j}=j_{\rm c}(e_{\rm i}/e_0)^{r-1} \bi{e}_{\rm i}/e_0,\label{power1}\eeq
where $r=1/p$ and $e_{\rm i}=|\bi{e}_{\rm i}|$. The critical current density $j_{\rm c}$ may depend
only on $(x_1,x_2)\in \Omega$ (the Bean model for an inhomogeneous film)
or also on the normal to the film component of the magnetic field (the Kim model).
It is convenient to assume that $\Omega$ is simply connected. If it contains holes, these can
simply be filled in, with the sheet critical current density
in the holes set to be zero or very small.

In the outer space $\omega:=\mathbb{R}^3\setminus \{\overline{\Omega}\times 0\}$
we have the Faraday and Ampere laws,
$$\mu_0\,\partial_t\bi{h}+\nabla\times \bi{e}=\bm{0},\qquad \nabla\times\bi{h}=\bm{0}$$
with $\bi{h}|_{t=0}=\bi{h}_0$ and $\bi{h}\rightarrow \bi{h}_{\rm e}(t)$ at infinity.
Here $\mu_0$ is the permeability of
vacuum, $\bi{e}$ and $\bi{h}$ are the electric and magnetic fields, respectively,
the given external magnetic field is uniform and normal to the film,
$\bi{h}_{\rm e}=(0,0,h_{\rm e}(t))$. We assume the initial magnetic field $\bi{h}_0$ has zero divergence,
$\nabla\cdot \bi{h}_0=0$, in $\omega$, its normal to the film component is continuous on $\{\Omega\times0\}$,
and $\bi{h}_0-\bi{h}_{\rm e}(0)=\Or(|x|^{-1})$ at infinity.

We now relate the exterior space and film problems, then use the magnetic scalar potential and
derive a  2D variational formulation,
written for the electric field $\bi{e}_{\rm i}$ and the jump of magnetic potential on the film,
which is convenient for the numerical approximation.

The jump of the tangential component of the magnetic field across
the cut $\{\Omega\times 0\}$
and the film current are related,
\beq \bi{j}=\bi{n}^+\times[\bi{h}],
\label{jnh}
\eeq
where $\bi{n}^+=(0,0,1)$. Here and below, $[\bi{f}]$ means the jump,
$\bi{f}|_{\Omega^+}-\bi{f}|_{\Omega^-}$, where $\Omega^{\pm}=\Omega\times
\{\pm 0\}$ are the two sides of $\{\Omega\times 0\}$.
Although the electric field $\bi{e}$ is not uniquely determined in $\omega$ by this model,
its tangential component on the film, $\bi{e}_{\tau}$, is and has to be continuous:
\beq \bi{e}_{\tau}|_{\Omega^+}=\bi{e}_{\tau}|_{\Omega^-}=\bi{e}_{\rm i}.
\label{econt}
\eeq
Since in the outer space  $\nabla\times\bi{h}=\nabla\times (\bi{h}-\bi{h}_{\rm e})=\bm{0}$
and $\Omega$ is
assumed to be simply connected, there exists a magnetic scalar potential $w(x,t)$ such that
$\bi{h}-\bi{h}_{\rm e}=-\nabla w$.
Furthermore, since $\nabla\cdot (\bi{h}-\bi{h}_{\rm e})=\nabla\cdot \bi{h}=0$,
the scalar potential is a harmonic function in $\omega$ for any $t$,
\beq \Delta w=0.
\label{Delw}
\eeq
Integrating the Faraday law in time we obtain
\beq \mu_0\,(\bi{h}-\bi{h}_0)+\nabla\times \bi{U}=\bm{0},
\label{iFLt}
\eeq
where $\bi{U}:=\int_0^t\bi{e}\,dt'$ has the continuous tangential component
$\bi{U}_{\tau}|_{\Omega^+}=\bi{U}_{\tau}|_{\Omega^-}=\bi{U}_{\rm i}:=\int_0^t\bi{e}_{\rm i}\, dt'$.
Noting that the normal component of magnetic field is continuous on the film and also that
$\bi{n}^+\!\cdot\nabla\times \bi{U}=\mbox{Curl}\,\bi{U}_{\rm i}$,
where $\mbox{Curl}\,  \bi{f}=\partial_{x_1}f_{2}- \partial_{x_2}f_{1}$,
we obtain that
 $$\mu_0\,\bi{n}^+\!\cdot(\bi{h}-\bi{h}_0)+\mbox{Curl}\,   \bi{U}_{\rm i}=0.$$
Substituting $\bi{h}=\bi{h}_{\rm e}-\nabla w$, we finalize our choice of the scalar potential $w$
as the solution to the following exterior problem:
$$\Delta w=0\qquad  \mbox{in}\ \omega,$$
\beq \frac{\partial w}{\partial \bi{n}^+}=\frac{1}{\mu_0}\,\mbox{Curl}\,
\left(\int_0^t\bi{e}_{\rm i}\,dt'\right)+{\cal H}\qquad \mbox{on}\ \Omega^+\ \mbox{and}\
\Omega^-,\label{exter}\eeq
$$w =\Or(|x|^{-1}) \qquad \mbox{as }|x| \rightarrow \infty,$$
with ${\cal H}=\bi{n}^+ \cdot\,(\bi{h}_{\rm e}-\bi{h}_0)$.

We set ${g}=[w]$ and note that if ${g}=0$ on the domain boundary $\partial\Omega$
and is sufficiently regular (belongs to the space $S_0=H^{1/2}_{00}(\Omega)$,
see \cite{BP00,BPmath}),
the unique solution to the following problem,
\begin{eqnarray*}
&\Delta w=0\qquad  \mbox{in } \omega,\\
&[w]={g},\qquad \left[\frac{\partial w}{\partial \bi{n}^+}\right]=0,\\
&w =\Or(|x|^{-1}) \qquad \mbox{as }|x| \rightarrow \infty
\end{eqnarray*}
is the double
layer potential (\cite{Nedelec}, Ch. 3, \S3.3 in the case of a closed surface $\Omega$,
and \cite{BPmath} for the present choice of $\Omega$)
$$w(x)=\frac{1}{4\pi}\int_{\Omega}
{g}(y)\,\frac{\partial}{\partial
\bi{n}^+_y}\!\left(\frac{1}{|x-y|}\right)dy \qquad \mbox{for } x \in \omega,$$
where $\partial/\partial \bi{n}^+_y=\bi{n}^+\cdot\nabla_y$.
The normal derivative of
this function, $\partial w/\partial \bi{n}^+$, is continuous across the
cut $\{\Omega\times 0\}$ and
satisfies the variational equation
\beq\int_{\Omega}\frac{\partial w}{\partial
\bi{n}^+}\,\psi \, dx =-a({g},\psi)\label{dwdn}\eeq
for any test function $\psi\in S_0$. Here the bilinear form
\begin{eqnarray}a({g},\psi)&=
\frac{1}{4\pi}\int_{\Omega}\int_{\Omega}
\frac{{\mbox{\bf Curl}}\,{g}(x)\cdot
{\mbox{\bf Curl}}\,\psi(y)}{|x-y|}\,dx\,dy\nonumber \\ &\equiv
\frac{1}{4\pi}\int_{\Omega}\int_{\Omega}
\frac{{\mbox{Grad}}\,{g}(x)\cdot {\mbox{Grad}}\,\psi(y)}{|x-y|}
\,dx\,dy\label{a_form}
\end{eqnarray} is symmetric, and $\mbox{\bf Curl}\,\phi=(\partial_{x_2}\phi,
-\partial_{x_1}\phi)$
and ${\rm Grad}\,\phi=(\partial_{x_1}\phi,\partial_{x_2}\phi)$ are 2D operators.
We note that $\frac{1}{2}a({g},{g})$ is the energy of the magnetic field
induced by the film current
\begin{eqnarray*}\bi{j}&=\bi{n}^+ \times[\bi{h}]=\bi{n}^+\times[\bi{h}_{\rm e}-\nabla w]\\&
=\mbox{\bf Curl}\,[w]=\mbox{\bf Curl}\,{g}.\end{eqnarray*}
Substituting the normal derivative of $w$ from (\ref{exter})
into the variational equation (\ref{dwdn}) we obtain
\beq a({g},\psi) +\frac{1}{\mu_0}
\left( \mbox{Curl} \left(\int_0^t
\bi{e}_{\rm i}\,dt' \right) ,\psi\right)_{\Omega} =
-\left({\cal H},\psi\right)_{\Omega}\label{i_form}\eeq
for any $\psi\in S_0$; here  $(u,v)_{\Omega}=\int_{\Omega}u\,v\,dx$
is the inner product (or duality pairing) of two functions on $\Omega$.
Differentiating with respect to time, we arrive at a more convenient form of this equation,
\beq a(\partial_t{g},\psi) +\frac{1}{\mu_0}
\left( \mbox{Curl}\,
\bi{e}_{\rm i} ,\psi\right)_{\Omega} =
-\left(\partial_th_{\rm e},\psi\right)_{\Omega}
\label{one0}\eeq
for any $\psi \in S_0$,
with ${g}|_{t=0}={g}_0$ determined by (\ref{i_form}) as
\beq
a({g}_0,\psi)=-(\,(\bi{h}_{\rm e}(0)-\bi{h}_0)\cdot\bi{n}^+,\psi)_\Omega
\label{inid}
\eeq for any $\psi \in S_0$.

Finally, since $\bi{j}=\mbox{\bf Curl}\,{g}$, we  rewrite the current
voltage relation (\ref{power1}) as
\beq \mbox{\bf Curl}\,{g}= j_{\rm c}({e}_{\rm i}/e_0)^{r-1} \bi{e}_{\rm i}/e_0\label{two0}\eeq
and arrive at the mixed variational formulation (\ref{one0})--(\ref{two0})
of the magnetization problem
written for two variables, $\bi{e}_{\rm i}$ and ${g}$, defined on $\Omega$ for any $t>0$.

It is convenient to use dimensionless variables, assuming
\begin{eqnarray*}&x=\frac{x'}{L},\ t= \frac{t'}{t_0},\
\bi{e}_{\rm i}=\frac{\bi{e}_{\rm i}'}{e_0},\\& \bi{j}=\frac{\bi{j}'}{j_{\rm c0}},\
\bi{h}=\frac{\bi{h}'}{j_{\rm c0}},\ {g}=\frac{{g}'}{j_{\rm c0}L},
\end{eqnarray*}
where $'$ denotes dimensional physical quantities, $2L$ is the length of the
projection of $\Omega$ onto the $x_1$-axis,  the
time scale $t_0= \mu_0j_{\rm c0}L/e_0$, and $j_{\rm c0}$ is a characteristic value of
the sheet critical current density.
For homogeneous films with the field independent critical density $j_{\rm c}$ we choose
in our simulations $j_{\rm c}/j_{\rm c0}=1$. 
If this density depends on the normal to the film component
of the magnetic field,
$j_{\rm c}=j_{\rm c}(h_3)$ on $\{\Omega\times 0\}$, one can take $j_{\rm c0}=j_{\rm c}(0)$.
The dimensionless form of the equations (\ref{one0}), (\ref{two0}) is
\beq a(\partial_t{g},\psi) +
\left( \mbox{Curl}\,
\bi{e}_{\rm i} ,\psi\right)_{\Omega} =
-\left(\partial_th_{\rm e},\psi\right)_{\Omega}
\label{one}\eeq
for any $\psi \in S_0$, and
\beq \mbox{\bf Curl}\,{g}= \frac{j_{\rm c}}{j_{\rm c0}}e_{\rm i}^{r-1} \bi{e}_{\rm i}.\label{two}\eeq
Computing the normal to the film magnetic field component
$h_3$ is needed in problems with field dependent critical sheet current densities,
and also for the comparison of numerical simulation results to magneto-optical imaging.
Noting that  $h_3-h_{\rm e}=-\partial w/\partial \bi{n}^+$ on $\{\Omega\times 0\}$,
we can use (\ref{dwdn}) for determining the magnetic field component $h_3$ from the equation
\beq \left(h_3-h_{\rm e},\psi\right)_{\Omega}=a({g},\psi)\label{h3}\eeq
for all $\psi\in S_0$. Alternatively, the explicit expression for $\partial w/\partial
\bi{n}^+$ in (\ref{exter}) yields, in dimensionless variables,
\beq h_3=h_{03}-{\rm Curl}\,\left(\int_0^t\bi{e}_{\rm i}\,dt'\right).
\label{h3div}\eeq
Yet another possibility \cite{P98} is to express the normal magnetic field component
via the potential jump (magnetization function) ${g}$ using the Biot-Savart law,
\begin{eqnarray} h_3(x,t)&=h_{\rm e}(t)+\bi{n}^+\cdot\frac{1}{4\pi}\int_{\Omega}\nabla_y
\left(\frac{1}{|x-y|}\right)\times\bi{j}(y,t)\,dy\nonumber \\
&=h_{\rm e}(t)-\frac{1}{4\pi}\int_{\Omega}{\rm Grad}_{\,y} \left(\frac{1}{|x-y|}\right)
\cdot{\rm Grad}_{\,y}\,{g}(y,t)\,dy.\label{h398}\end{eqnarray}
These three approaches are further discussed in Sec. \ref{NS}.

\section{Numerical scheme}\label{NS}
It is important to approximate the electric field $\bi{e}_{\rm i}$ in problem
(\ref{one})--(\ref{two}) using curl conforming finite elements.
In 2D problems, a simple change of variables leads to a formulation
where  curls are replaced by divergences;  the divergence conforming
Raviart-Thomas elements (see below) are an appropriate choice for such formulations.

Let us substitute $\bi{e}_{\rm i}=R\,\bi{v}$, where  $R$ is the rotation matrix
$$ \left( \begin{array}{rr}0 &\ 1 \\
-1 &\ 0 \end{array} \right).$$
Taking into account that $|\bi{v}|=|\bi{e}_{i}|$, $\mbox{Curl}\, R = -{\rm Div}$
and $R^T\,\mbox{\bf Curl}={\rm Grad}$,  we rewrite (\ref{one})--(\ref{two})
as
\begin{eqnarray}
&a(\partial_t{g},\psi)-\left({\rm Div}\,\bi{v},\psi\right)_{\Omega}
=-\left(\partial_th_{\rm e},\psi\right)_{\Omega},
\label{one_v}\\
& {\rm Grad}\,{g}=\frac{j_{\rm c}}{j_{\rm c0}}|\bi{v}|^{r-1} \bi{v}\label{two_v}\end{eqnarray}
for any $\psi \in S_0$.
Here $\mbox{Div}\,\bi{v}=\partial_{x_1}v_1+\partial_{x_2}v_2$ is the 2D divergence.
Multiplying equation (\ref{two_v}) by a vector test function $\bm{\eta}$ and using Green's formula,
we rewrite this equation as
\beq  j_{\rm c0}^{-1}(j_{\rm c}|\bi{v}|^{r-1}\bi{v},\bm{\eta})_{\Omega}+({g},{\rm Div}\,\bm{\eta})_{\Omega}=0.
\label{two_v1}\eeq
Equation (\ref{h3div}) should also be rewritten:
\beq h_3=h_{03}+{\rm Div}\left(\int_0^t\bi{v}\,dt'\right).\label{h3divv}\eeq
We approximate $\Omega$ by a polygonal domain $\Omega^h$. Let ${\cal T}^h$
be a regular partitioning of $\Omega^h$  into triangles
$\kappa$
and $h=\max_{\kappa \in {\cal T}^h}{\rm diam}(\kappa)$ be their maximal size.
Here vertices of ${\cal T}^h$ lying on $\partial \Omega^h$, the boundary of $\Omega^h$,
also lie on $\partial \Omega$. If $\Omega$ contains subdomains with different critical
current density values, the mesh is fitted in a similar way to the subdomain boundaries.
By ${\cal N}^h$ and ${\cal E}^h$ we denote the sets of nodes and edges of this triangulation,
respectively, with ${\cal N}_{\rm i}^h$ being the subset of the internal and ${\cal N}_{\rm b}^h$
of the boundary nodes. Below,
$|{\cal X}|$ will denote the number of elements in the set ${\cal X}$.

Let $S_0^h$ be the space of continuous functions, linear on each triangle, and zero in
the boundary nodes ${\cal N}^h_{\rm b}$. We define also the finite dimensional space of vectorial
functions ${\cal V}^h$, linear on each triangle,
$\bi{v}^h|_{\kappa}=\bi{a}_{\kappa}+b_{\kappa}(x_1,x_2)$, $\bi{a}_{\kappa}\in \mathbb{{R}}^2,$
$b_{\kappa}\in \mathbb{{R}}^1$ and such that the normal component of $\bi{v}^h$
is continuous across any edge separating two adjacent triangles in  ${\cal T}^h$.
This is the space of divergence conforming Raviart-Thomas elements of the lowest order;
see \cite{Carst} for a detailed description of the edge related basis for ${\cal V}^h$.

In addition, let
$0= t_0 < t_1 < \ldots < t_{N-1} < t_N = T$ be a
partitioning of $[0,T]$
into possibly variable time steps $\tau_n = t_n -
t_{n-1}$, $n=1\to N$. 
\\ Our approximation of the problem (\ref{one_v}), (\ref{two_v1}) is:

Given ${G}^0 \in S^h_0$, for $n = 1 \rightarrow N$,
find ${G}^n \in S^h_0$
and $\bi{V}^n \in
{\cal V}^h$ such that
\begin{eqnarray}&a^h({G}^n,\psi^h) - {\tau_n}
\left({\rm Div}\,\bi{V}^n,
\psi^h\right)_{\Omega^h} =
a^h({G}^{n-1},\psi^h)
-\left(h_{\rm e}^n - h_{\rm e}^{n-1},\psi^h\right)_{\Omega^h},
\label{Qaeh}\\
&j_{\rm c0}^{-1}\left(j_{\rm c}\,|\bi{V}^n|^{r-1}\,\bi{V}^n,
\bm{\eta}^h\right)^h+
\left({G}^n,{\rm Div}\,\bm{\eta}^h\right)_{\Omega^h}= 0 \label{Qbeh}
\end{eqnarray}
for all $\psi^h \in S^h_0$ and $\bm{\eta}^h \in {\cal
V}^h\,$. Here $h_{\rm e}^n$ denotes $h_{\rm e}(t_n)$, $a^h(.,.)$ is defined
by (\ref{a_form}) with $\Omega$ replaced by $\Omega^h$, and $(\bi{f},\bi{u})^h
=\sum_{\kappa\in{\cal T}^h}(\bi{f},\bi{u})^h_{\kappa}$ averages the integrand $\bi{f}\cdot
\bi{u}$ over each triangle $\kappa$
at its vertices:
$$ (\bi{f},\bi{u})^h_{\kappa}=\frac{1}{3}\, |\kappa|\, \sum_{m=1}^3 \,
\bi{f}(P_m^{\kappa})\cdot\bi{u}(P_m^{\kappa}),$$
where $\{P_m^{\kappa}\}_{m=1}^3$ are the vertices of
triangle $\kappa$ and $|\kappa|$ its area.

Furthermore, ${G}^0 \in S^h_0$ solves the corresponding approximation
of (\ref{inid}). We note that it is not necessary to solve explicitly for ${G}^0$
as we can just replace the first term on the right-hand side of (\ref{Qaeh}) for $n=1$
by $({h}_{03}-{h}_{\rm e}(0),\psi^h)_{\Omega^h}$.

It is easy to show the existence and uniqueness of a solution to the
nonlinear algebraic system (\ref{Qaeh})--(\ref{Qbeh}),
see \cite{BPmath}. To solve this system at each time level,
we set $\bi{V}^{n,0}=\bi{V}^{n-1}$, denote $|b|_{\epsilon}=\sqrt{|b|^2+\epsilon^2}$
and approximate $|\bi{V}^n|^{r-1}\,\bi{V}^n$ at the $j^{\rm th}$ iteration by
$$|\bi{V}^{n,j-1}|^{r-1}\,\bi{V}^{n,j-1}+(|\bi{V}^{n,j-1}|_{\epsilon})^{r-1}\,(\bi{V}^{n,j}-\bi{V}^{n,j-1});$$
and find ${G}^{n,j} \in S^h_0$
and $\bi{V}^{n,j} \in
{\cal V}^h$ such that
\begin{eqnarray}&   a^h({G}^{n,j},\psi^h) -{\tau_n}
\left(\mbox{Div}\,\bi{V}^{n,j},
\,\psi^h\right)_{\Omega^h}\nonumber\\ &
\ \ \ = a^h({G}^{n-1},\psi^h)
-\left(h_{\rm e}^n - h_{\rm e}^{n-1})
,\psi^h\right)_{\Omega^h}
\label{Qaehj}\end{eqnarray}
\begin{eqnarray}&  j_{\rm c0}^{-1}\left(j_{\rm c}\,|\bi{V}^{n,j-1}|^{r-1}_{\epsilon}\,
\bi{V}^{n,j},\bm{\eta}^h \right)^h+\left(
{G}^{n,j}, \mbox{Div}\,\bm{\eta}^h\right)_{\Omega^h}\nonumber\\ &
\ \ \
=j_{\rm c0}^{-1}\left(j_{\rm c}\left( \,|\bi{V}^{n,j-1}|_{\epsilon}^{r-1}-|\bi{V}^{n,j-1}|^{r-1}\,\right)
\bi{V}^{n,j-1}\, ,\bm{\eta}^h \right)^h
\label{Qbehj}
\end{eqnarray}
for all $\psi^h \in S^h_0$ and $\bm{\eta}^h \in {\cal
V}^h\,$.
At each iteration, we need to solve the following linear system
\begin{eqnarray*}&
A \,\underline{{G}}^{j} - \tau_n B \,\underline{V}^{j} = \underline{d}\,,\\
&B^T\,\underline{{G}}^{j}+M^{j-1} \,\underline{V}^{j}
=\underline{f}^{j-1}
\end{eqnarray*}
to determine ${G}^{n,j}=\sum_{k=1\rightarrow |{\cal N}^h_{\rm i}|}\underline{{G}}^j_k\,\psi_k$
and $\bi{V}^{n,j}=\sum_{k=1\rightarrow |{\cal E}^h|}\underline{V}^j_k\,\bm{\eta}_k;$
here $\{\psi_k\}$ and $\{\bm{\eta}_k\}$ are the standard bases for $S^h_0$ and ${\cal V}^h$,
respectively, and the time index $n$ is omitted for simplicity.

Here $A$ is a symmetric positive definite full $|{\cal N}^h_{\rm i}|
\times
|{\cal N}^h_{\rm i}|$ matrix with elements $A_{k,l}=a^h(\psi_k,\psi_l)$;
$M^{j-1}$ is  a symmetric positive definite sparse
$|{\cal E}^h| \times |{\cal E}^h|$ matrix with elements
\beq M^{j-1}_{k,l}=j_{\rm c0}^{-1}\,
(j_{\rm c}\,|\bi{V}^{n,j-1}|_{\epsilon}^{r-1}\bm{\eta}_k,\bm{\eta}_l)^h;\label{M_el}\eeq
and $B$ is a sparse $|{\cal N}^h_{\rm i}| \times|{\cal E}^h|$
matrix with elements $B_{k,l}=(\psi_k,{\rm Div}\,\bm{\eta}_l)_{\Omega^h}$.

We found that convergence of these iterations can be accelerated by supplementing them
with an over-relaxation,   i.e., by recalculating $\bi{V}^{n,j}$
as $\alpha \bi{V}^{n,j}+(1-\alpha)\bi{V}^{n,j-1}$ with $\alpha>1$
after each iteration. In all the examples below we chose
$\epsilon=10^{-6}$ and $\alpha=1.2$.

We also note that only the sparse matrix $M^{j-1}$  must be recalculated at each iteration.
The full matrix $A$ is calculated only once for the chosen finite element mesh.
Since gradients of the basis functions $\psi_k$ are constant on each triangle,
to calculate these matrix elements one should find, see (\ref{a_form}),
the double surface integrals
$$\int_{\kappa_l}\int_{\kappa_m}
\frac{dx\,dy}{|x-y|}
$$
for every pair of triangles $\kappa_l,\,\kappa_m\in {\cal T}^h$.
We note that some of these integrals  are singular. To accurately approximate this matrix,
we followed the approach in the
appendix of \cite{SP10}; in particular, we used the exact analytical value \cite{Arcioni}
for the most singular cases $l=m$.

To compare simulation results with the magneto-optical measurements of the
normal to the film component of the magnetic field
$h_3$, an approximation $H_3$  to this component can be computed
at the inner mesh nodes ${\cal N}^h_{\rm i}$ using a
discretized form of the variational equation (\ref{h3}),
\beq \underline{H}_3^n=h_{\rm e}(t_n)\,\underline{\it 1}+\Lambda^{-1}\,A\,\underline{{G}}^n,\label{H3}\eeq
where $\underline{\it 1}$ is the $|{\cal N}^h_{\rm i}|\times 1$ vector $(1,1,...,1)^T$
and $\Lambda$ is the diagonal $|{\cal N}^h_{\rm i}|
\times
|{\cal N}^h_{\rm i}|$ matrix with $\Lambda_{k,k}=\int_{\Omega}\psi_k\, dx$. Note that,
due to the infinitely thin film approximation employed in our model,
this magnetic field component becomes
infinite on the domain boundary; see, e.g., the thin strip solution \cite{BrandtIndenbomForkl}.

If the critical current density depends on the normal to the
film magnetic field component,
$j_{\rm c}=j_{\rm c}(h_3)$, it is necessary to substitute
$j_{\rm c}$ by $j_{\rm c}(H_3^{n,j-1})$ in (\ref{M_el}) and to update
the approximation $H_3$
after each iteration. The inner node $H_3$ values (\ref{H3}) are not convenient
for approximating $j_{\rm c}$ in all triangles, as is needed in (\ref{M_el}).

   The piecewise constant approximation $\widetilde{\underline{H}}_3$
resulting from a discretization of  (\ref{h3divv}), can be written as
\beq \widetilde{\underline{H}}_3^{\,j}=\widetilde{\underline{H}}_{03}
+C\,(\underline{U}^{n-1}+\tau_n\,\underline{V}^j),\eeq
where $\widetilde{H}_{03}$ is a piecewise constant approximation of $h_{03}$, $C$
is the sparse $|{\cal T}^h|\times|{\cal E}^h|$ matrix with elements $C_{m,l}
={\rm Div}(\bm{\eta}_l)|_{\kappa_m}$,
and $\underline{U}^{n-1}$ denotes the coefficients of
$\bi{U}^{n-1}=\sum_{m=1}^{n-1} \tau_m\,\bi{V}^m$
for the standard basis of the Raviart-Thomas space ${\cal V}^h$.
We found, however, that such an approximation, based on the integrated over time electric
field, can be inaccurate, especially in and around the film holes (in which the electric field
remains undetermined in our model).

A better piecewise constant approximation was obtained using a discretized form of
(\ref{h398}): we set
$\widetilde{\underline{H}}_3^{\,j}|_{\kappa}:=\widetilde{{H}}_3^{\,j}(p_{\kappa})$, where
$p_{\kappa}$ is the center of triangle ${\kappa}$ and
\beq\widetilde{{H}}_3^{\,j}(p_{\kappa})=h_{\rm e}^n-\frac{1}{4\pi}
\sum_{\kappa'\in{\cal T}^h}\left( {\rm Grad}\,
{G}^{\,j}\right)|_{\kappa'}\cdot\oint_{\partial \kappa'}\frac{\bm{\nu}_{\kappa'}}{|p_{\kappa}-s|}\,ds.
\label{Biot}\eeq
Here $\partial\kappa'$ is the boundary of $\kappa'$, $\bm{\nu}_{\kappa'}$
is the unit outward normal to this boundary,
and the integral over each side of triangle $\kappa'$ is computed
numerically (as in \cite{P98} we simply used Simpson's quadrature rule).

\section{Simulation results}
The simulations have been performed in Matlab R2011a (64 bit) on a PC with
Intel Core i5-2400 3.10Hz processor and 4Gb RAM. All film magnetization problems
were solved for a growing external field $h_{\rm e}(t)=t$ and a zero field initial state.

First, to test our method, we solved numerically the thin disk problem. Let $\Omega$
 be a circle of radius one. For the Bean critical state model the exact distribution of
 the sheet current density $\bi{j}=j(\rho,t)\,\widehat{\bm{\phi}}$ is known \cite{MikhKuz,ClemSanchez}.
 Here $\widehat{\bm{\phi}}$ is the unit azimuthal vector in polar coordinates $(\rho,\phi)$.
 In our dimensionless variables,
 $$j(\rho,t)=\left\{\begin{array}{lr}
-1 & a(t)\leq \rho\leq 1,\\
-\frac{2}{\pi}\arctan\left\{\rho\sqrt{\frac{1-a^2(t)}{a^2(t)-\rho^2}}\right\} & 0\leq \rho<a(t),\end{array}\right.
$$
where $a(t)=1/\cosh\left(2h_{\rm e}(t)\right).$ The normal to the film component of the magnetic
field can be found by means of a 1D numerical integration using the equation
$$h_3(\rho,t)=h_{\rm e}(t)+\frac{1}{2\pi}\int_0^1G(\rho,\rho')\,j(\rho',t)\,d\rho',$$
where
$G(\rho,\rho')=K(k)/(\rho+\rho')-E(k)/(\rho-\rho')$,
$k=2\sqrt{\rho\,\rho'}/(\rho+\rho')$ and $K$ and $E$
are complete elliptic integrals of the first and second kind.
Furthermore, the electric field $\bi{e}=e(\rho,t)\,\widehat{\bm{\phi}}$, 
where ${\rho}^{-1}\partial_\rho(\rho\,e)=-\partial_t h_3,\quad e|_{\rho=0}=0.$
Approximating $\partial_t h_3$ by $\{h_3(\rho,t_n)-h_3(\rho,t_{n-1})\}/\tau_n$ and
integrating numerically, we calculate an approximation to the
electric field distribution averaged over the time interval
$(t_{n-1},t_n)$.

\begin{figure}[h!]
\begin{center}
\includegraphics[width=12cm]{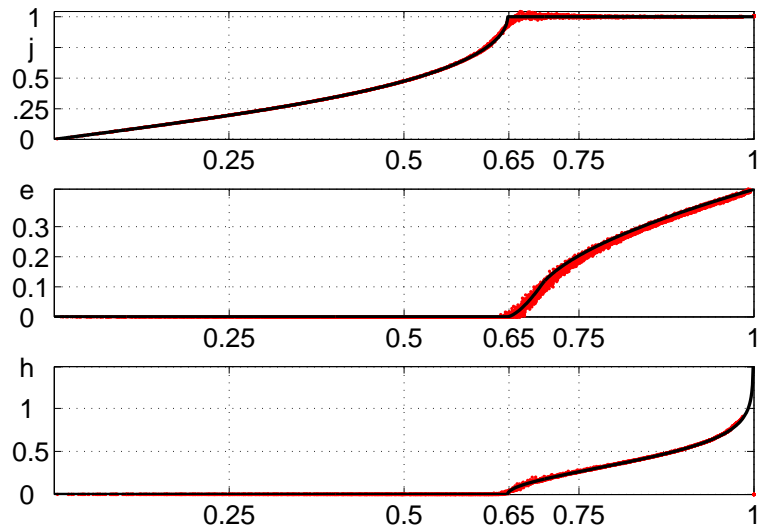}
\end{center}
\caption{Thin disk in the perpendicular field,  $h_{\rm e}(t)=t$. The Bean model solution
\cite{MikhKuz,ClemSanchez} (black line) and the numerical solution (red dots) obtained
with $p=1000$, $h=0.03$. Shown for $t=0.5$:
top -- the modulus of the current density $j$; middle -- the modulus of the electric field
$e_{\rm i}$; bottom -- the normal component of the magnetic field, $h_3$.}
\label{Fig1}
\end{figure}

To compare with this semi-analytical solution of the Bean model,
we set $p=1000$ in the power law model,  and used our
numerical algorithm  (\ref{Qaeh})--(\ref{Qbeh})
with $r=1/p$. 
Our numerical experiments confirmed that for a monotonically  growing external field
the current density and the magnetic field can be computed in one time step
without any accuracy loss. The electric field $\bi{e}_{\rm i}$ is, however,
determined by time derivatives of the external magnetic field and the magnetization function;
in the discretized formulation (\ref{Qaeh})--(\ref{Qbeh}) the field $\bi{E}^n=R\,\bi{V}^n$
can be considered as an approximation to the field $\bi{e}_{\rm i}$ at some time moment
in the interval $(t_{n-1},t_n)$ or to the average field in this time interval.
Hence, our numerical strategy was to make a large time step $\tau_1$
followed by a much smaller step $\tau_2$ to obtain accurate approximations to all variables,
including the electric field, at time $t=\tau_1+\tau_2$.
The simulation results for $t=0.5$ (Fig. \ref{Fig1}) were obtained with
$\tau_1=0.45,\tau_2=0.05$ and two finite element meshes, with 4200 elements ($h=0.05$)
and 12000 elements ($h=0.03$). Solution for these two meshes took, respectively,
5 and 57 minutes of
CPU time, not including the time for computing the full matrix $A$.
Comparing to the solution of the Bean model described above,
we found that the critical current zone was $0.65\leq \rho\leq 1$,
where $a(0.5)=1/\coth(1)\approx 0.65$,
and the relative errors for the current density, the electric field,
and the normal magnetic field component were, respectively,
1\%, 4.9\%, and 2.5\% for the crude mesh and
0.6\%, 3.1\%, 1.4\% for the fine mesh. Here the electric field in the
semi-analytical solution for the Bean model was calculated
using the $h_3$ distributions at the same two time moments, $t_1=0.45$, $t_2=0.5$.
At each time level, the approximate current density was computed  as ${\bf Curl}\, {G}^n$,
constant in each
triangle. However, for the comparison we used the node values  calculated, at each node,
as the weighted-by-areas mean of the values in triangles to which the node belongs;
such averaging increased the accuracy. We note also that the magnetic field $h_3$
was determined using equation (\ref{H3}). Hence, the field was found and compared to the
exact solution at the internal nodes only (at the boundary nodes the exact field is infinite).

In the next two examples we also assumed $p=1000$, so the numerical solutions
obtained should be close to solutions to the Bean model; the magnetic field
was computed using the discretized Biot-Savart formula (\ref{Biot}).

\begin{figure}[h!]
\begin{center}
\includegraphics[width=12cm]{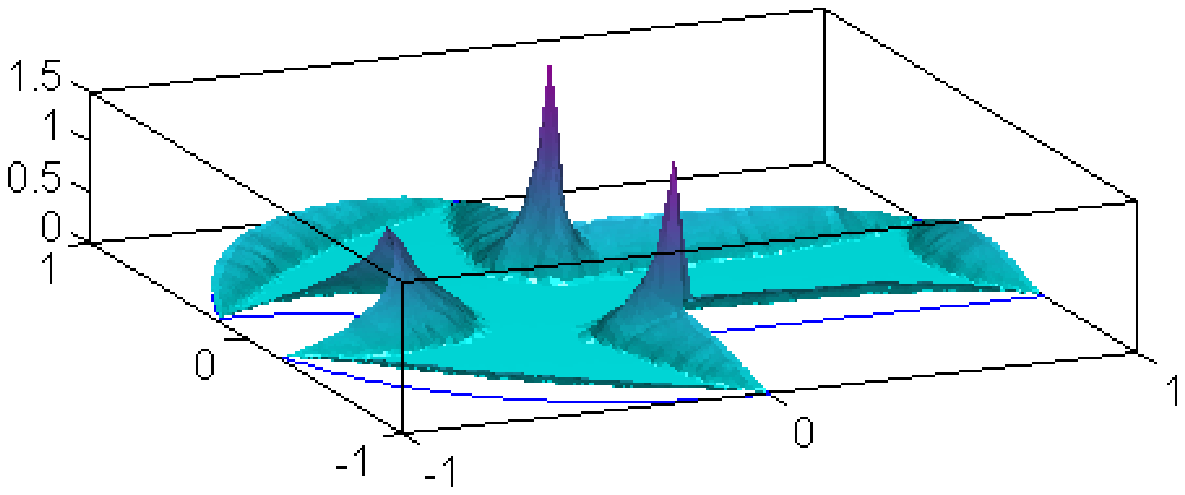}\\
\includegraphics[width=6cm]{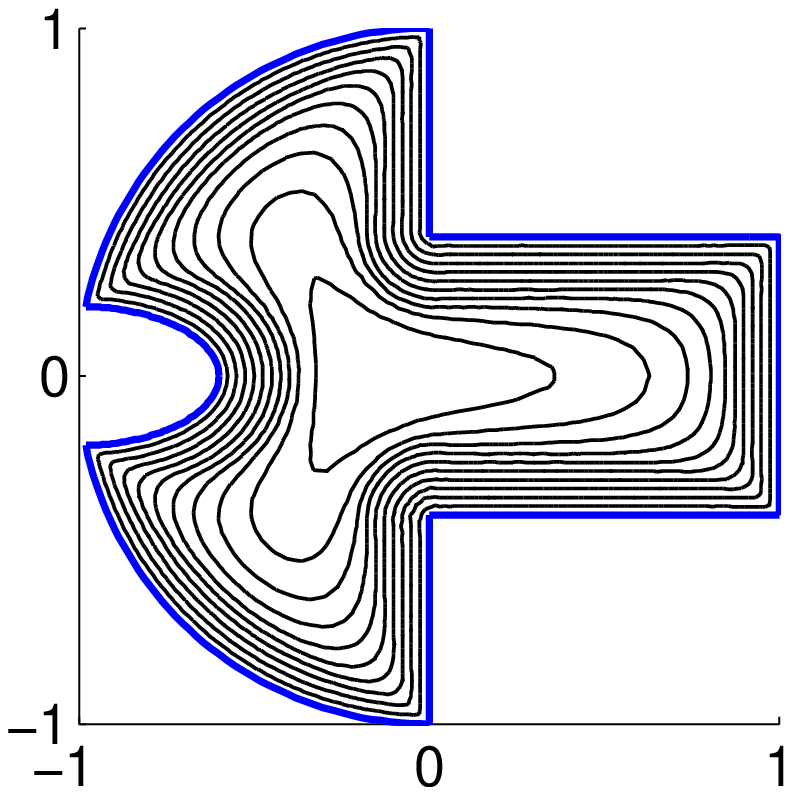}\hspace{.6cm}\includegraphics[width=6cm]{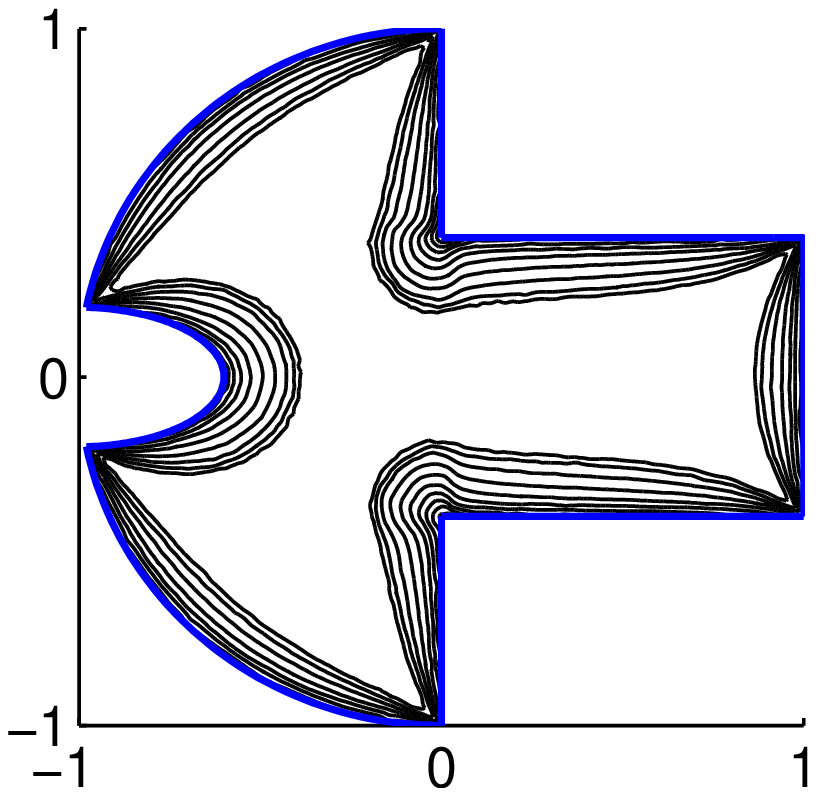}
\end{center}
\caption{A film with corners and boundary indentation, $h_{\rm e}(t)=t$;
numerical solution for $p=1000$. Shown for $t=0.4$:
top -- the modulus of the electric field $e_{\rm i}$, bottom -- current streamlines (left) and
levels of the normal to the film magnetic field component $h_3$ (right).
The mesh (about 9000 
triangles) was refined near the film boundary.}
\label{Fig_cross_ind}
\end{figure}

The electric field is known to be strong near the film boundary indentations and, especially,
in the vicinity of concave film corners (see Fig. \ref{Fig_cross_ind}).
Although similar problems
have been solved by other authors before,
this was done for $p=9$ and $p=19$ in \cite{SchusterB96}
and \cite{VSGJ07,VSGJ08}, respectively  (and also for the critical state models in \cite{P98},
but there without computing the electric field).
Here we took time steps $0.3+0.08+0.02=0.4$.

\begin{figure}[h!]
\begin{center}
\begin{minipage}[h]{7.6cm}\includegraphics[width=7.6cm,height=12cm]{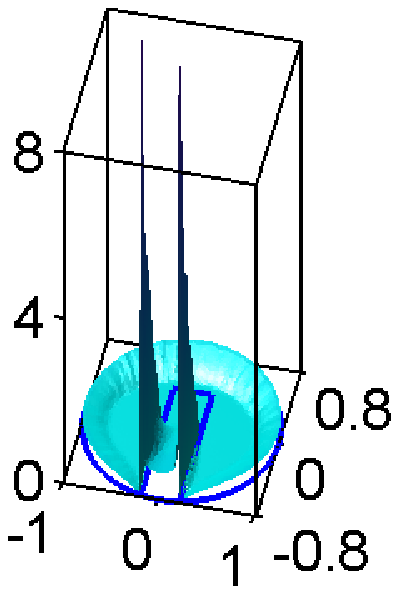}\end{minipage}
\begin{minipage}[h]{7.5cm} \includegraphics[width=7.5cm,height=6cm]{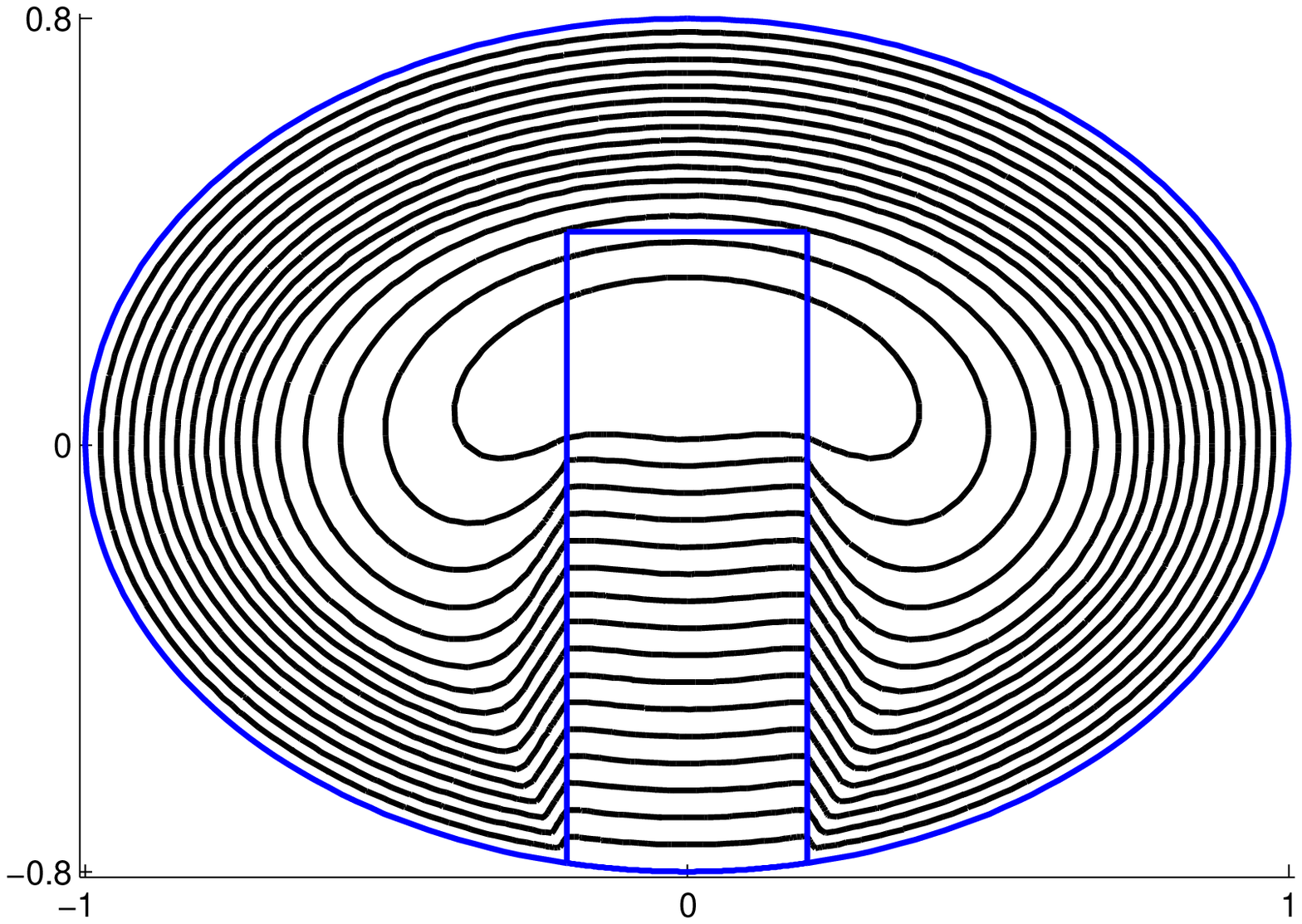}\\
\includegraphics[width=7.5cm,height=6cm]{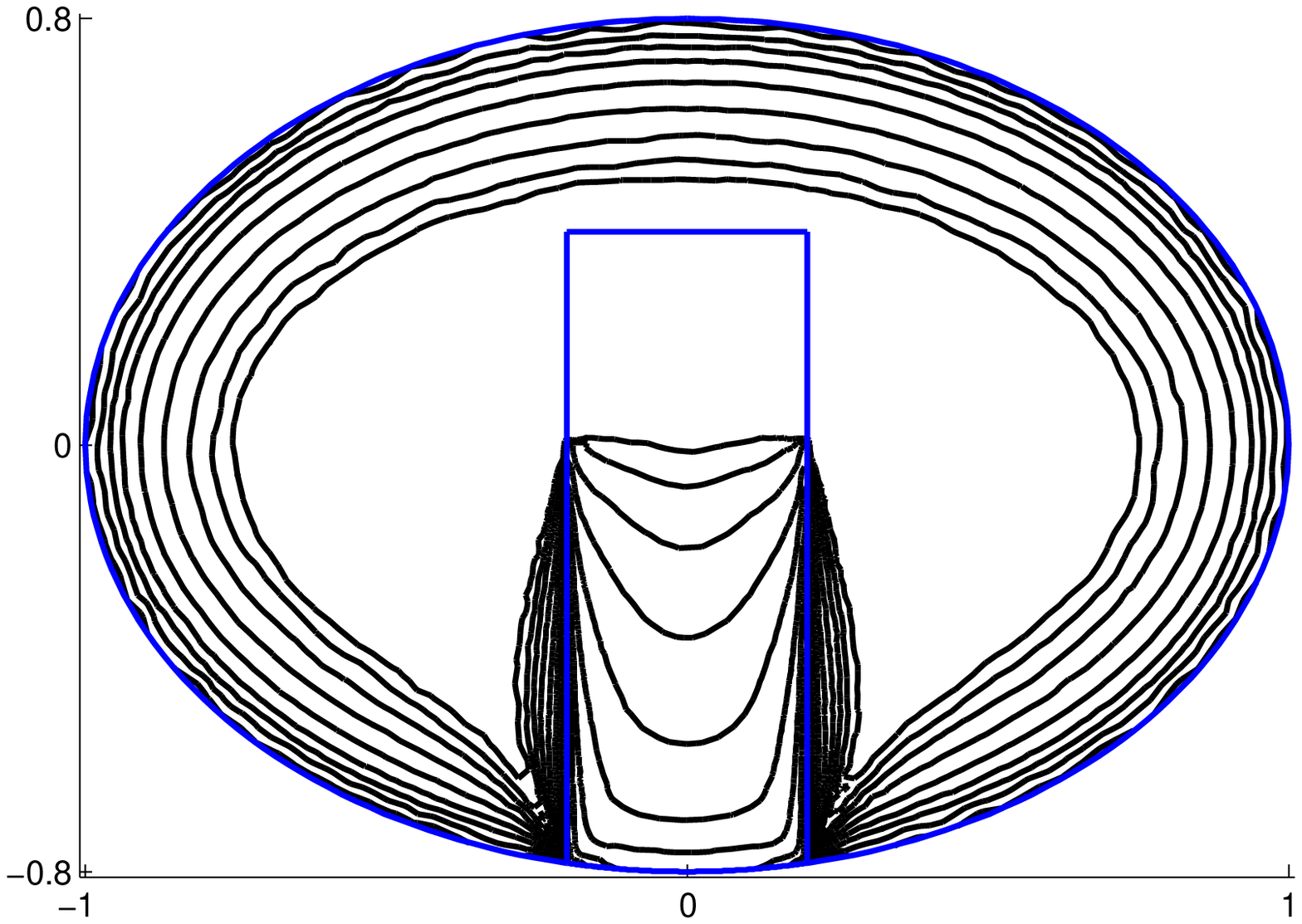}
\end{minipage}
\end{center}
\caption{Inhomogeneous film of elliptic shape in a growing external field; $p=1000$.
Sheet critical current density $j_{\rm c}/j_{\rm c0}=0.5$ in the rectangle
and $j_{\rm c}/j_{\rm c0}=1$
outside of it. The finite element mesh contained 10,600 
triangles and was refined
near the boundary between the two regions (the blue line). Shown for $t=0.5$:
left -- the modulus of the electric field $e_{\rm i}$; right --
current streamlines (top) and levels of the normal to the film magnetic field
component $h_3$ (bottom).}
\label{Fig_inh}
\end{figure}

In inhomogeneous films the electric field near the boundaries between regions of different
critical current densities can be orders of magnitude higher
than in other parts of the film
(see Fig. \ref{Fig_inh}).
Here $j_{\rm c}/j_{\rm c0}=0.5$ inside the rectangle, and $j_{\rm c}/j_{\rm c0}=1$ outside.
We took time steps $0.2+0.2+0.1=0.5$.
The magnetic flux penetrates deeper into the lower critical current
density area.

\begin{figure}[h!]
\begin{center}
\begin{minipage}[h]{7.6cm}\includegraphics[width=7.6cm,height=12cm]{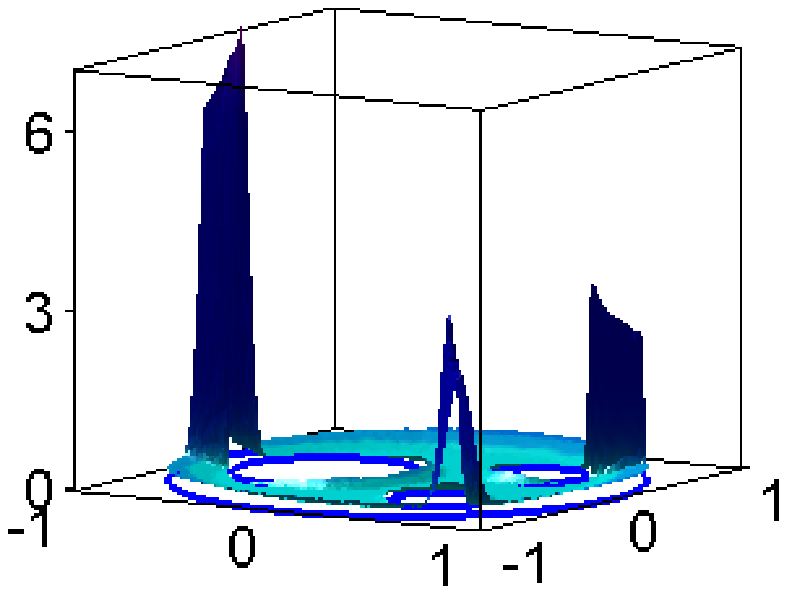}\end{minipage}
\begin{minipage}[h]{7.5cm} \includegraphics[height=6cm]{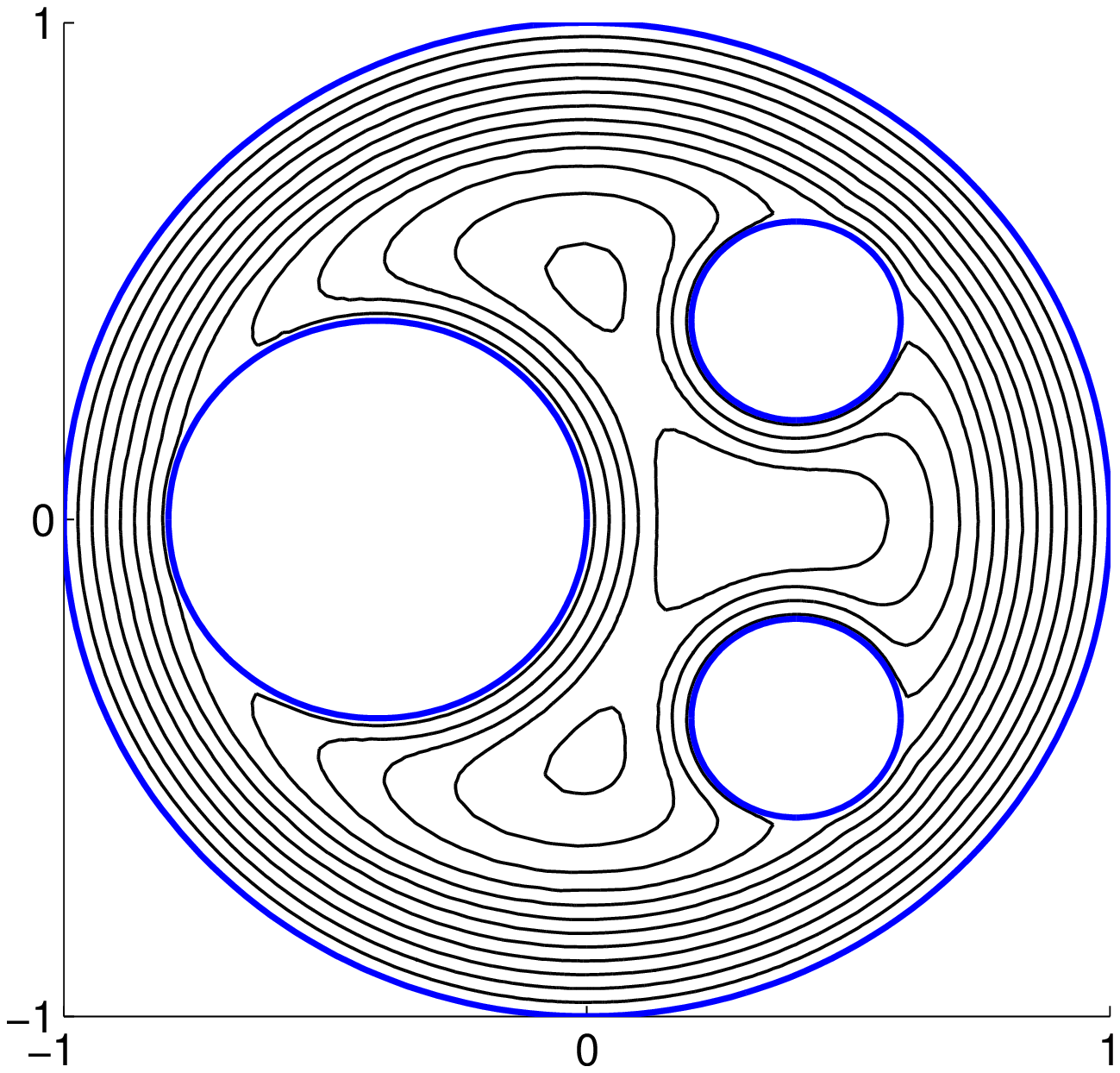}\\
\includegraphics[height=6cm]{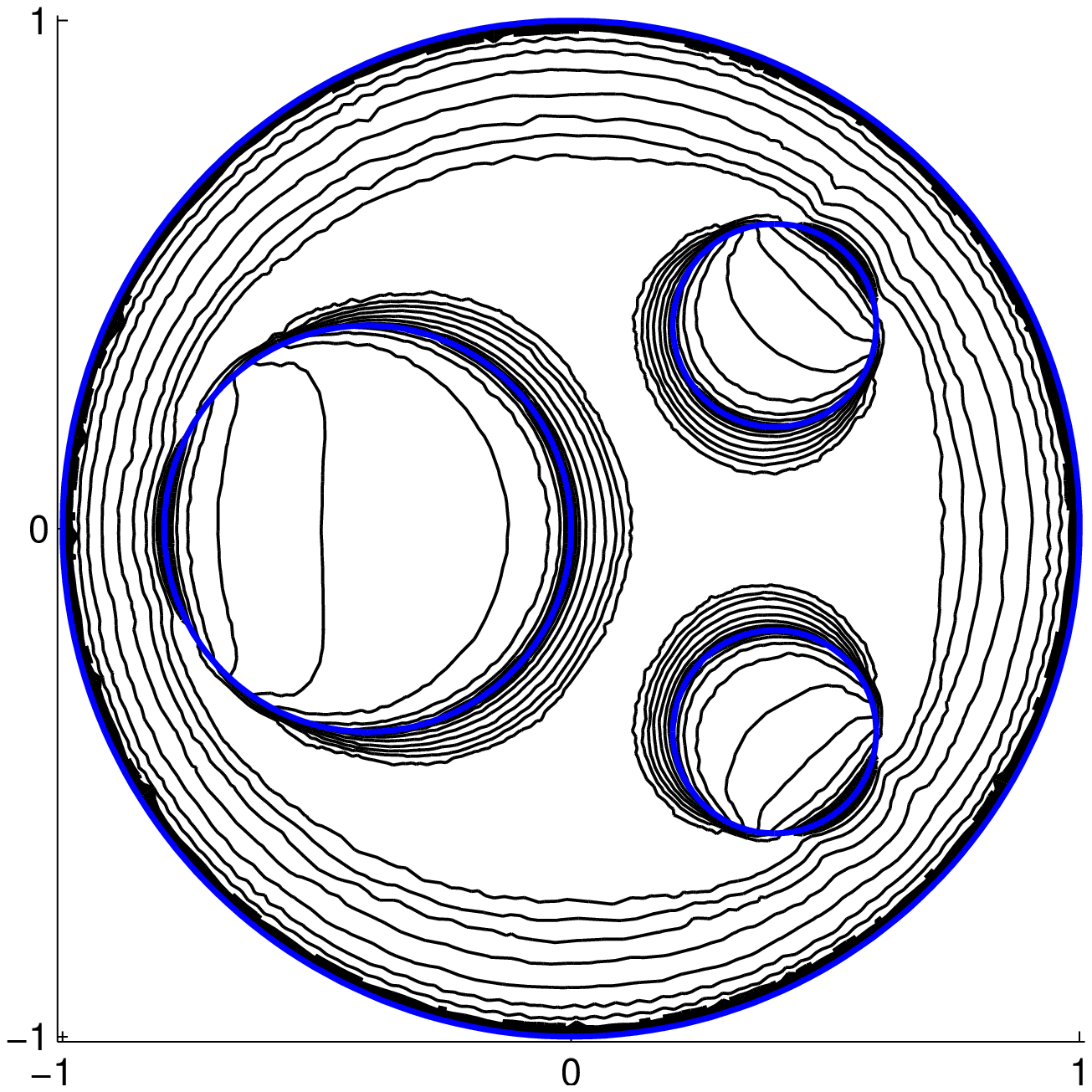}
\end{minipage}
\end{center}
\caption{Circular film with three holes in a growing external field; $p=100$.  The finite
element mesh contained 10,400 
triangles and was refined near the domain and hole
boundaries (blue lines). Shown for $t=0.5$: left -- the modulus of the electric field
$e_{\rm i}$; right -- current streamlines (top) and levels of the normal to the film
magnetic field component $h_3$ (bottom).}
\label{Fig_3holes}
\end{figure}

To solve a problem with a multiply connected film (Fig. \ref{Fig_3holes}) we filled the holes
in and set $j_{\rm c}/j_{\rm c0}=0.002$ there, while keeping  $j_{\rm c}/j_{\rm c0}=1$
in the film itself; we recall the electric field in
the holes is not determined. This example was solved for $p=100$
with time steps $0.47+0.03=0.5$.
A strong electric field is
generated along the paths of flux penetration into the holes. For $p=19$ such problems
were solved by a different method in \cite{VSGJ08}.

\section{Conclusion}
Existing numerical methods for thin film magnetization problems in type-II superconductivity
are based on formulations written for one main variable: the magnetization function.
The sheet current density, determined numerically as the curl of this function,
is prone to numerical inaccuracy. The inaccuracy, usually tolerable in the current density
itself, inhibits evaluation of the electric field by substituting this density into
a power current-voltage relation if the power is high.
For critical state models such an approach for computing the electric field
is not applicable.

The new variational formulation of thin film magnetization problems proposed in this work
is written for two variables, the electric field and the magnetization function.
The formulation serves as a basis for the approximation and computation of all variables of
interest: the sheet current density and both the electric and magnetic fields.
Our numerical algorithm remains accurate for any value of the power in the power
law current-voltage relation. For high powers we obtain a good approximation to
the solution of
the Bean model. Evaluation of the local heat dissipation distribution in a film for both
the power law and critical state models becomes straightforward.

In this paper, we presented numerical simulation results for isotropic models
with field independent critical sheet current density.
However, our approach can be generalized to thin film problems with field-dependent \cite{Kim}
and anisotropic \cite{Schuster97} sheet critical current densities.

\section*{\bf Acknowledgement} L.P.\ appreciates helpful discussions with V.\ Sokolovsky.

\end{document}